\documentclass{article}
\usepackage{e}
\usepackage{graphicx}
\newcommand\beq{\begin{eqnarray}}
\newcommand\eeq{\end{eqnarray}}
\newcommand\bq{\begin{equation}}
\newcommand\eq{\end{equation}}
\newcommand\qom{\frac{q}{M}}
\newcommand\qove{\frac{q}{E_{\nu}}}
\newcommand\me{\frac{m_{\nu}}{E_{\nu}}}
\newcommand\qovm{\frac{q}{2M}}
\newcommand\fip{\frac{\mid \phi_{\mu}(0) \mid^{2}}{4\pi}}

\newcommand\scr{\mid C_{S}^{R} \mid^{2}}

\newcommand\invcl{\mid C_{V}^{L} + 2 M
g_{M}\mid^{2}}
\newcommand\inacl{\mid C_{A}^{L} +
  m_{\mu}\qovm g_{P}\mid^{2}}
\newcommand\invlsr{(C_{V}^{L}+ 2 M
g_{M})C_{S}^{R*}}
\newcommand\invlal{(C_{V}^{L}+ 2 M
g_{M})(C_{A}^{L*} +
  m_{\mu}\qovm g_{P}^{*})}

\newcommand\inalpr{(C_{A}^{L} +
  m_{\mu}\qovm g_{P})C_{P}^{R*}}
\newcommand\inaltr{(C_{A}^{L} +
  m_{\mu}\qovm g_{P})C_{T}^{R*}}

\newcommand\invltr{(C_{V}^{L} + 2 M
g_{M})C_{T}^{R*}}

\newcommand\tcr{\mid C_{T}^{R} \mid^{2}}
\newcommand\srtr{C_{S}^{R}C_{T}^{R*}}
\newcommand\trpr{C_{T}^{R}C_{P}^{R*}}

\newcommand\pmo{{\bf |P_{\mu}|}}
\newcommand\pq{{\bf (\hat{P}_{\mu}\cdot \hat{q})}}

\begin{document}
\branch{A}
\DOI{123}                       
\idline{A}{1, 1--11}{1}         
\editorial{}{}{}{}              
\title{Right-handed Dirac Neutrinos in $\nu e^{-}$ Scattering and Azimuthal Asymmetry
in Recoil Electron Event Rates}
\author{}
\author{S. Ciechanowicz\inst{1}\and M. Misiaszek\inst{2}
\and W. Sobk\'ow\inst{1}}
%
\institute{Institute of Theoretical Physics, University of
Wroc\l{}aw, Pl. M. Borna 9,\\ PL-50-204~Wroc\l{}aw, Poland
\\
e-mail: ciechano@rose.ift.uni.wroc.pl, sobkow@rose.ift.uni.wroc.pl
\and  M. Smoluchowski Institute of
Physics, Jagiellonian University, ul. Reymonta 4,\\ PL-30-059 Krak\'ow, Poland\\
email: misiaszek@zefir.if.uj.edu.pl}
%
\maketitle
\begin{abstract}
In this paper a scenario with the participation  of the exotic
scalar S, tensor T and pseudoscalar P couplings of the
right-handed neutrinos in addition to the standard vector V,
axial A couplings of the left-handed neutrinos in the low-energy
$(\nu_{\mu}e^{-})$ and $(\nu_{e}e^{-})$ scattering  processes is
considered. Neutrinos are assumed to be massive Dirac fermions
and to be polarized. Both reactions are studied at the level of
the four-fermion point interaction. The main goal is to show that
the physical consequence of the presence of the right-handed
neutrinos is an appearance of the azimuthal asymmetry in the
angular distribution of the recoil electrons caused by the
non-vanishing interference terms between the standard and exotic
couplings, proportional to the transverse neutrino polarization
vector. The upper limits on the expected effect of this asymmetry
for the low-energy neutrinos $(E_{\nu} < 1 \, MeV)$ are found. We
also show that if the neutrino helicity rotation $(\nu_{L}
\rightarrow \nu_{R})$ in the solar magnetic field takes place, the
similar effect of the azimuthal asymmetry of the recoil electrons
scattered by the solar neutrinos should be observed. This effect
would also come from the interference terms between the standard
$(V, A)_{L}$ and exotic $(S, T, P)_{R}$ couplings. New-type
neutrino detectors with good angular resolution could search for
the azimuthal asymmetry in event number.

\end{abstract}
\section{Introduction}
The standard vector-axial $(V-A)$ structure  of the neutral and
charged  weak interactions describes only what  has been measured
so far. We mean here  the measurement of the electron helicity
\cite{Bob}, the indirect measurement of the neutrino helicity
\cite{Gold}, the asymmetry in the distribution of the electrons
from $\beta$-decay \cite{CWu} and the experiment with muon decay
\cite{Gar} which confirmed parity violation \cite{Lee}. Feynman,
Gell-Mann and independently Sudarshan, Marshak \cite{Gell}
established that only left-handed vector $ V$, axial $ A$
couplings can take part in weak interactions because this yields
the maximum symmetry breaking under space inversion, under charge
conjugation; the two-component neutrino theory of negative
helicity;  the conservation of the combined symmetry $ CP$ and of
the lepton number. It means  that produced neutrinos
(antineutrinos) in $V-A$ interaction can only be left-handed
(right-handed). However Wu \cite{SWu} pointed out that  exotic
scalar  S, tensor T and pseudoscalar  P weak interactions may be
responsible for the negative electron helicity observed in
$\beta$-decay. It would suggest that the generated neutrinos
(antineutrinos) in the  $(S, T, P)$ interactions may also be {\it
right-handed} ({\it left-handed}). The experimental precision of
present measurements still does not rule out the possible
participation of the exotic $(S, T, P)$ couplings of the
right-handed neutrinos beyond the the Standard Model (SM)
\cite{Glashow,Wein,Salam}.
\par So Sromicki at the PSI \cite{Sromicki} searched for $T$-odd
transverse electron polarization in  $^{8}Li$ $\beta$-decay.
Armbruster {\it et al.} \cite{Armbr} measured the energy spectrum
of electron-neutrinos $\nu_{e}$ from $\mu$-decay at rest in the
KARMEN experiment using the reaction
$^{12}C(\nu_{e},e^{-})^{12}N_{g.s.}$. They gave the upper limit on
the magnitude of interference between scalar $S$ and tensor $T$
couplings. Shimizu {\it et al.}  \cite{Shimizu} determined the
ratio of the strengths of scalar and tensor couplings to the
standard vector coupling in $K^{+}\rightarrow
\pi^{0}+e^{+}+\nu_{e}$ decay at rest assuming the only left-handed
neutrinos for all interactions. Bodek {\it et al.}  at the PSI
\cite{Bodek}  looked for the evidence of the violation of time
reversal invariance measuring $T$-odd transverse positron
polarization in $\mu^{+}$-decay. They also admitted the presence
of the only left-handed neutrinos produced in the scalar
interaction. The emiT collaboration presented  new limits on the
time reversal invariance violating D coefficient using the
polarized neutron beta-decay \cite{Neutron}.  Presently at PSI,
the experiment with the decay of polarized neutrons is prepared
to search for the time reversal violating effects. A non-zero
value of the T-odd transverse component of the electron
polarization  would be a signal of the violation  of this symmetry
\cite{Barnett}.  The transverse electron polarization for the
electrons in the decay of polarized muons was  calculated by
Shekhter and Okun \cite{Okun} in 1958. The recent results
presented by the DELPHI Collaboration \cite{Delphi} concerning
the measurement of the Michel parameters and the neutrino
helicity in $\tau$ lepton decays  still admit the deviation from
the standard $V-A$ structure of the charged current weak
interaction.
 \par  New
high-precision low-energy tests of the  Lorentz structure using
the electron-neutrinos coming from the strong and polarized
low-energy artificial neutrino source or from  the Sun  would be
sensitive to the  effects caused by the interference terms
between the standard $(V, A)_{L}$ couplings of the left-handed
neutrinos and exotic $(S, T, P)_{R}$ couplings of the right-handed
neutrinos in the neutrino-electron scattering.
\par So far the neutrino-electron scattering was proposed to
measure the azimuthal asymmetry in the recoil electron event rates
produced by the non-zero neutrino magnetic moments
 in the case of the
 solar neutrinos \cite{Barbieri,Pastor}. This asymmetry is caused by the
 non-vanishing
interference between the weak and electro-magnetic interaction
amplitudes, proportional to $\mu_{\nu}$, and depends on the
azimuthal angle between the transverse component of the neutrino
polarization and the momentum of the outgoing recoil electron.
 \par The first concept of the use of
the artificial neutrino source comes from Alvarez who proposed a
$^{65}Zn$ \cite{Alvarez}. The $^{51}Cr$ and $^{37}Ar$ neutrino
sources were proposed by Raghavan \cite{Raghavan} in 1978 and
Haxton \cite{Haxton} in 1988, respectively. The idea of  using
the artificial neutrino source (reactor neutrinos) to search for
the neutrino magnetic moments was first proposed by Vogel and
Engel \cite{Vogel}. The strong $^{51}Cr$ source was used for the
calibration of the GALLEX neutrino experiment \cite{Gallex}.
Miranda {\it et al.} \cite{Miran} proposed the use of the
$^{51}Cr$ source to probe the gauge structure of the electroweak
interaction. Currently at Gran Sasso, the Borexino neutrino
experiment \cite{Borexino} with the unpolarized $^{51}Cr$ source
is designed to search for the neutrino magnetic moment. There are
also proposed the other experiments to test the non-standard
properties of neutrinos, in which both the recoil electron
scattering angle and the azimuthal angle would be measured with
good precision: the Hellaz \cite{Hellaz}, the Heron \cite{Heron}.
\par In this paper, we show  that there is the other possible
scenario of the appearance of the azimuthal asymmetry in the
differential  cross section for the neutrino-electron scattering.
The participation of the exotic $(S, T, P )_{R}$ couplings in
addition to the standard $(V, A)_{L}$ couplings  can generate the
azimuthal asymmetry in the event number because in the final
state (after scattering) all the neutrinos are left-handed, and
the interference terms between the standard and exotic couplings
do not depend on the neutrino mass. The main goal is to find the
upper limits on the expected magnitude of the azimuthal asymmetry
in the angular distribution of the recoil electrons for the
low-energy $(\nu_{\mu}e^{-})$ and $(\nu_{e}e^{-})$ scattering
processes $(E_{\nu}=0.746 \, MeV, E_{\nu}=0.863 \, MeV)$, using
the current limits on the non-standard couplings \cite{Data}. This
paper is also a generalization of the considerations made in the
\cite{Sobkow}.  The obtained results are analyzed in the context
of the future low-energy high-precision neutrino experiments.
\par In
our considerations the system of natural units with $\hbar=c=1$,
Dirac-Pauli representation of the $\gamma$-matrices and the $(+,
-, -, -)$ metric are used \cite{Mandl}.
\section{Muon capture by proton as production process of neutrinos}
\begin{figure}
\includegraphics[width=9cm,angle=0]{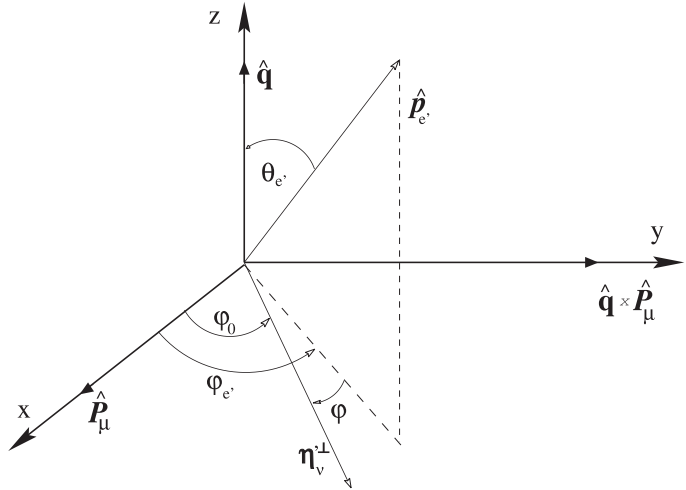}
\caption{The production plane and reaction plane for $\nu_\mu$
neutrinos with the transverse neutrino polarization vector
$\mbox{\boldmath $\eta_{\nu}^{' \perp}$}$.}
 \label{wsp}
\end{figure}

To show how the transverse components of the neutrino
polarization, both $T$-even and $T$-odd, may appear, we use the
reaction of the muon capture by proton $(\mu^{-} + p \rightarrow
n + \nu_{\mu})$ as a production process of muon-neutrinos.  The
production plane is spanned by the direction of the initial muon
polarization ${\bf {\hat P}_{\mu}}$ and of the outgoing neutrino
momentum ${\bf \hat{q}}$, Fig. \ref{wsp}. ${\bf {\hat P}_{\mu}}$
and ${\bf \hat{q}}$ are assumed to be perpendicular to each other
because this leads to the unique conclusions as to the possible
presence of the right-handed neutrinos. Govaerts and
Lucio-Martinez \cite{Govaerts} considered the nuclear muon
capture on the proton and $^{3} He $ both within and beyond SM
admitting the general Lorentz invariant four-fermion contact
interaction and assuming the Dirac massless neutrino. However,
they did not calculate the neutrino observables.
One assumes that the outgoing neutrino flux is a mixture of the
left-handed neutrinos produced in the standard $V-A$ charged weak
interaction and the right-handed ones produced in the exotic
scalar $S$, pseudoscalar $P$, tensor $T$ charged weak
interactions. In this scenario the interacting muon is always
left-handed. The complex fundamental coupling constants are
denoted as $C_{V}^{L}, C_{A}^{L}$ and $C_{S}^{R}, C_{P}^{R},
C_{T}^{R}$ respectively to the outgoing neutrino L- and
R-chirality: \beq {\cal M}_{\mu^-} & = & (C_{V}^{L}+ 2 M
g_{M})({\overline u}_{\nu}
    \gamma_{\lambda}(1 - \gamma_{5})u_{\mu})
    ({\overline u}_{n}\gamma^{\lambda}u_{p})\\
  && \mbox{}+ (C_{A}^{L}+
  m_{\mu}\qovm g_{P})({\overline u}_{\nu}
    i\gamma_{5}\gamma_{\lambda}(1 - \gamma_{5})u_{\mu})
    ({\overline u}_{n}i\gamma^{5}\gamma^{\lambda}u_{p})\nonumber\\
&& \mbox{} + C_{S}^{R}({\overline u}_{\nu}(1 -
\gamma_{5})u_{\mu})({\overline u}_{n} u_{p})  +
C_{P}^{R}({\overline u}_{\nu}\gamma_{5}(1-\gamma_{5})u_{\mu})
    ({\overline u}_{n}\gamma_{5}u_{p})\nonumber\\
   && \mbox{} + C_{T}^{R}({\overline u}_{\nu}\sigma_{\lambda\,\rho}
    (1-\gamma_{5})u_{\mu})({\overline u}_{n}\sigma^{\lambda\,\rho} u_{p}),
\nonumber\eeq where  $g_{M}, g_{P}$ - the induced couplings of
the left-handed neutrinos, i.e. the weak magnetism and induced
pseudoscalar, respectively;
 $m_{\mu}, q, E_{\nu}, m_{\nu}, M$ - the muon mass, the value of the neutrino momentum, its
 energy, its mass and the nucleon mass; $u_{p}, {\overline u}_{n}$ - the Dirac bispinors
 of initial proton and final neutron; $u_{\mu}, {\overline u}_{\nu}$ - the Dirac bispinors
 of initial muon and final neutrino.
\par The received formulas for the neutrino observables, in the case
of non-vanishing neutrino mass $(m_{\nu}\not =0)$, when the
induced couplings are enclosed and  ${\bf \hat{P}_{\mu}}$, ${\bf
\hat{q}}$ are perpendicular to each other
$(\pq =0)$, are as follows:\\
  T-even transverse component of the neutrino polarization:
 \beq\label{neut3}
    \lefteqn{{\bf <S_{\nu}\cdot{\hat P}_{\mu}}>_f
 = \null\fip\pmo Re\{(1 + \qove\qovm)\invlsr} \\
 & & \mbox{} + \qove\qovm\inalpr
+ 2\qove\qovm\invltr \nonumber \\
& & \mbox{} + 2(1 + \qove\qovm)\inaltr \nonumber\\
  &  & \null + \frac{1}{2} \me( \invcl - \inacl  + \scr - 2\tcr)\}. \nonumber
  \eeq
T-odd transverse component of the neutrino polarization:
  \beq\label{neut2}
    \lefteqn{{\bf <S_{\nu}\cdot({\hat P}_{\mu}\times\hat{q})>}_f
   = \null\fip\pmo Im\{-(\qove + \qovm)\invlsr}   \\
  &&\null - \qovm\inalpr - 2\qovm\invltr  \nonumber \\
&  & \null - 2(\qove + \qovm)\inaltr + 2\me\qovm(\srtr -
\trpr)\}, \nonumber \eeq where  ${\bf S_{\nu}}$ - the operator of
the  neutrino spin; $\pmo$ - the value of the muon polarization in
$1s$ state; $\phi_{\mu}(0)$ - the value of the large radial
component of the muon Dirac bispinor for $r=0$. The above
neutrino observables are calculated with the use of the density
matrix of the final state. \par   It can be noticed that the
neutrino observables consist only of the interference terms
between the standard $(V, A )_{L}$ couplings of the left-handed
neutrinos
 and exotic  $(S, T, P)_{R}$ ones of the right-handed neutrinos in the limit of vanishing neutrino
 mass.
 It can be understood as the interference between the neutrino
waves of negative and positive chirality. There is no
 contribution to these observables from the SM in which neutrinos
 are only left-handed and  massless.  The
mass terms in the above neutrino observables give a very small
contribution in relation to the main one coming from the
interference terms  and they are  neglected in the
considerations.  If one assumes the production of the only
left-handed neutrinos in all the interactions, i.e. both for the
standard $V-A$  and $(S, T, P)$ interactions, there is no
interference between the standard $C_{V,A}^{L}$ and
$C_{S,T,P}^{L}$ couplings in the limit of vanishing neutrino
mass. We see that the induced couplings enter additively to the
fundamental $C_{V,A}^{L}$ couplings and they are  omitted in the
considerations because their presence does not change
qualitatively the conclusions concerning the transverse neutrino
polarization. \par All the fundamental coupling constants
$C_{V,A}^{L}, C_{S, T, P }^{R}$ can be expressed by the couplings
$g^{\gamma}_{\epsilon \mu}$ for the normal and inverse muon decay
\cite{Data}, assuming the universality of weak interactions.
Here, $\gamma= S, V, T$ indicates a scalar, vector, tensor
interaction; $\epsilon, \mu=L, R$ indicate the chirality of the
electron or muon and the neutrino chiralities are uniquely
determined for given $\gamma, \epsilon, \mu$. We get the
following relations: \begin{equation} C_{V}^{L} = A(g_{LL}^{V} +
g_{RL}^{V}),\;  -C_{A}^{L} = A(g_{LL}^{V} - g_{RL}^{V}),
\end{equation} \bq C_{S}^{R} = A(g_{LL}^{S} + g_{RL}^{S}), \;
-C_{P}^{R} = A(g_{LL}^{S} - g_{RL}^{S}),  \;  C_{T}^{R} =
A(g_{LL}^{T} + g_{RL}^{T}),  \eq where
$A\equiv(4G_{F}/\sqrt{2})cos\theta_{c}$, $G_{F}= 1.16639(1)\times
10^{-5}GeV^{-2}$  is the Fermi coupling constant \cite{Data},
$\theta_{c}$ is the Cabbibo angle. We calculate the lower limits
on the $C^{L}_{V,A}$ and upper limit on the $C_{S, T, P }^{R}$,
using the current data \cite{Data}: $|C_{V}^{L}|>0.850 \,A,
\;|C_{A}^{L}|>1.070 \,A, \;|C_{S}^{R}|<0.974 \,A,
\;|C_{P}^{R}|<0.126 \,A, \;|C_{T}^{R}|<0.122 \,A$. In this way,
we get the upper bound on the   magnitude  of the transverse
neutrino polarization vector proportional
 to the value of the muon polarization:
\beq
 |\mbox{\boldmath $\eta_{\nu }^{\perp}$}| & = & \frac{\sqrt{{\bf <
S_{\nu}\cdot({\hat P}_{\mu}\times {\hat q}) >}_{f}^{2} + {\bf <
S_{\nu}\cdot {\hat P}_{\mu}>}_{f}^{2}}}{s <\bf 1>_{f}}
 \leq 0.414 \,\pmo \\
  |\mbox{\boldmath $\eta_{\nu}^{' \perp}$}| &=& \frac{|\mbox{\boldmath
$\eta_{\nu }^{\perp}$}|}{\pmo}  \leq  0.414,
 \eeq
where $s$ is the neutrino spin $(s=1/2)$ and the probability of
muon capture $<\bf 1>_{f}$ is of the form:
 \beq
{\bf <1>_{f}}
& =  & \mbox{}  \fip \{(1 + 2 \qove \qovm )\invcl + \scr \\
&  &  \mbox{} + (3 + 2 \qove \qovm)\inacl +
(12+8\qove\qovm)\tcr \nonumber\\
&& \mbox{}  + 2 Re[\qove \qom(-\invlal \nonumber\\
 &  & \mbox{} + \trpr + \srtr) + \me( \invlsr \nonumber \\ && \mbox{} -
 6\inaltr)] \}. \nonumber \eeq
 The obtained limit on the
 $|\mbox{\boldmath $\eta_{\nu}^{\perp}$}|$ has to be divided by $\pmo$
 to have the physical value of
 the $|\mbox{\boldmath $\eta_{\nu }^{' \perp}$}|$ generated by the exotic $(S, T, P)$ interactions.
 It means that the  value of the longitudinal neutrino polarization is
equal to  $\mbox{\boldmath $\hat{\eta}_{\nu}$}\cdot\hat{\bf q} =
-0.910$.
The  formula for the T-even  longitudinal component of the
neutrino polarization is as follows:
\beq \lefteqn{{\bf <S_{\nu}\cdot{\hat q}>}_f
 = \null\fip\{-({3\over 2}\qove+\qovm)\inacl}\\
 & & \mbox{} - ({1\over 2}\qove + \qovm)\invcl
+ {1\over 2}\qove\scr + (6\qove + 8\qovm)\tcr       \nonumber\\
  &&\null+ 2Re[\qovm\invlal + \qovm\srtr + \qovm\trpr          \nonumber\\
  &&\null - \me\qovm({1\over 2}\inalpr+\inaltr \nonumber\\
 && \mbox{} + {1\over 2}\invlsr+\invltr)]\} \nonumber.
  \nonumber\eeq
We see that in the longitudinal neutrino polarization and the
probability of muon capture, the occurrence of the interference
terms between the standard $C_{V, A}^{L}$ couplings and exotic
$C_{S, T, P}^{R}$ ones depends explicitly on the neutrino mass.
The dependence on the neutrino mass causes the "conspiracy" of
the interference terms and makes the measurement of the relative
phase   between these couplings impossible because term
$(m_{\nu}/E_{\nu})(q/2M)$ is very small and  the standard $C_{V,
A}^{L}$ couplings of the left-handed neutrinos dominate in
agreement with the SM prediction. Therefore, the neutrino
observables in which such difficulties do not appear are
proposed. In this way, the conclusions as to the existence of the
right-handed neutrinos can depend on  the type of measured
observables. \par If $m_{\nu} \rightarrow 0, \; q/E_{\nu}
\rightarrow 1 $  and the neutrino mass terms vanish in all the
observables.
\section{Neutrino-electron scattering as detection process}
The produced mixture of the muon-neutrinos is detected in  the
neutral current weak interaction.  We assume that the incoming
left-handed neutrinos are detected in the $V-A$ neutral weak
interaction, while the initial right-handed ones are detected in
the  exotic scalar $S$, tensor $T$ and pseudoscalar P neutral
weak interactions. Then  in the final state all the neutrinos are
left-handed.
\par To describe $(\nu_{\mu }e^{-})$ scattering the following
observables are used: \mbox{\boldmath $\hat{\eta}_{\nu}$} - the
unit 3-vector of the initial neutrino polarization in its rest
frame, ${\bf q}$ - the incoming neutrino momentum, ${\bf p_{e'}}$
- the outgoing electron momentum. The coupling constants are
denoted as $g_{V}^{L}, g_{A}^{L}$ and $g_{S}^{R}$, $g_{T}^{R}$,
$g_{P}^{R}$ respectively to the incoming neutrino L- and
R-chirality: \beq { \cal M}_{\nu e} &=&
\frac{G_{F}}{\sqrt{2}}\{(\overline{u}_{e'}\gamma_{\alpha}(g_{V}^{L}
- g_{A}^{L}\gamma_{5})u_{e}) (\overline{u}_{\nu '}
\gamma^{\alpha}(1 - \gamma_{5})u_{\nu}) \\
& & \mbox{} +
\frac{1}{2}[(g_{S}^{R}(\overline{u}_{e'}u_{e})(\overline{u}_{\nu
'} (1 + \gamma_{5})u_{\nu}) +
g_{T}^{R}(\overline{u}_{e'}\sigma_{\alpha \beta }
u_{e})(\overline{u}_{\nu '} \sigma^{\alpha \beta }(1 +
\gamma_{5})u_{\nu}) \nonumber \\ & & \mbox{} +
(g_{P}^{R}(\overline{u}_{e'}\gamma_{5}u_{e})(\overline{u}_{\nu '}
\gamma_{5}(1 + \gamma_{5})u_{\nu})] \nonumber\},
 \eeq
 where $ u_{e}$ and  $\overline{u}_{e'}$
$(u_{\nu}\;$ and $\; \overline{u}_{\nu '})$ are the Dirac
bispinors of the initial and final electron (neutrino)
respectively.
\subsection{Laboratory differential cross section}
 The laboratory differential cross section for the
$\nu_{\mu}e^{-}$ scattering, in the limit of vanishing neutrino
mass, is of the form: \beq \label{przekr} \frac{d^{2} \sigma}{d y
d \phi_{e'}} &=& (\frac{d^{2} \sigma}{d y d \phi_{e'}})_{(V, A)}
+ (\frac{d^{2} \sigma}{d y d \phi_{e'}})_{(S, T, P)}
\\
&& \mbox{} + (\frac{d^{2} \sigma}{d y d \phi_{e'}})_{(V S)} +
(\frac{d^{2} \sigma}{d y d \phi_{e'}})_{(A T)}, \nonumber  \\
(\frac{d^{2} \sigma}{d y d \phi_{e'}})_{(V, A)} &=&
\frac{E_{\nu}m_{e}}{4\pi^2} \frac{G_{F}^{2}}{2} \{
(1-\mbox{\boldmath $\hat{\eta}_{\nu}$}\cdot\hat{\bf q}
)[(g_{V}^{L} +
g_{A}^{L})^{2} + (g_{V}^{L}- g_{A}^{L})^{2}(1-y)^{2}\\
&& \mbox{} - \frac{m_{e}y}{E_{\nu}}((g_{V}^{L})^{2} - (g_{A}^{L})^{2})]\}, \nonumber \\
(\frac{d^{2} \sigma}{d y d \phi_{e'}})_{(S, T, P)} &=& \mbox{}
\frac{E_{\nu}m_{e}}{4\pi^2} \frac{G_{F}^{2}}{2}(1+\mbox{\boldmath
$\hat{\eta}_{\nu}$}\cdot\hat{\bf
q})\{\frac{1}{8}y(y+2\frac{m_{e}}{E_{\nu}})
 |g_{S}^{R}|^{2} + \frac{1}{8}y^{2}|g_{P}^{R}|^{2}  \\ && \mbox{} +
  ((2-y)^2 -
\frac{m_{e}}{E_{\nu}}y)|{g_{T}^{R}}|^{2} +
y(y-2)\frac{1}{2}[Re(g_{S}^{R}g_{T}^{*R}) +
Re(g_{P}^{R}g_{T}^{*R})]\}, \nonumber\\
 (\frac{d^{2}\sigma}{d y d \phi_{e'}})_{(V S)} &=& \mbox{} \frac{E_{\nu}m_{e}}{4\pi^2}
\frac{G_{F}^{2}}{2}\{
\sqrt{y(y+2\frac{m_{e}}{E_{\nu}})}[-\mbox{\boldmath
$\hat{\eta}_{\nu}$}\cdot({\bf \hat{p}_{e'} \times
\hat{q}})Im(g_{V}^{L}g_{S}^{R*}) \\ && \mbox{} + (\mbox{\boldmath
$\hat{\eta}_{\nu}$}\cdot {\bf \hat{p}_{e'}})
Re(g_{V}^{L}g_{S}^{R*})] -
 y(1+\frac{m_{e}}{E_{\nu}}) (\mbox{\boldmath
$\hat{\eta}_{\nu}$}\cdot\hat{\bf q}) Re(g_{V}^{L}g_{S}^{R*})\},
\nonumber \\
(\frac{d^{2} \sigma}{d y d \phi_{e'}})_{(A T)} &=&
\frac{E_{\nu}m_{e}}{4\pi^2} \frac{G_{F}^{2}}{2}\{ 2
\sqrt{y(y+2\frac{m_{e}}{E_{\nu}})}[-\mbox{\boldmath
$\hat{\eta}_{\nu}$}\cdot({\bf \hat{p}_{e'} \times
\hat{q}})Im(g_{A}^{L}g_{T}^{R*}) \\ && \mbox{} + (\mbox{\boldmath
$\hat{\eta}_{\nu}$}\cdot {\bf \hat{p}_{e'}})
Re(g_{A}^{L}g_{T}^{R*})] - 2 y(1+\frac{m_{e}}{E_{\nu}})
(\mbox{\boldmath $\hat{\eta}_{\nu}$}\cdot\hat{\bf q})
Re(g_{A}^{L}g_{T}^{R*})\}, \nonumber \eeq
where:  \beq  y & \equiv &
\frac{T_{e}}{E_{\nu}}=\frac{m_{e}}{E_{\nu}}\frac{2cos^{2}\theta_{e'}}
{(1+\frac{m_{e}}{E_{\nu}})^{2}-cos^{2}\theta_{e'}}, \eeq where
$y$ - the ratio of the kinetic energy of the recoil electron
$T_{e}$  to the incoming neutrino energy $E_{\nu}$, $\theta_{e'}$
- the angle between the direction of the outgoing electron
momentum  $ \hat{\bf p}_{e'}$  and the direction  of the incoming
neutrino momentum $\hat{\bf q}$ (recoil electron scattering
angle), $m_{e}$ - the electron mass, $ \mbox{\boldmath
$\hat{\eta}_{\nu}$}\cdot\hat{\bf q}$ - the longitudinal
polarization of the incoming neutrino, $\phi_{e'}$ - the angle
between the production plane and the reaction plane spanned by
the $ \hat{\bf p}_{e'}$ and $ \hat{\bf q}$, Fig. \ref{wsp}.  All
the calculations are made with the Michel-Wightman density matrix
\cite{Michel} for the polarized incoming neutrinos in the limit
of vanishing neutrino mass (see Appendix). The interference terms
between the standard and exotic couplings, Eqs. (14, 15), include
only the contributions from the transverse components of the
initial neutrino polarization, both $T$-even and $T$-odd: \beq
\label{inter} (\frac{d^{2} \sigma}{d y d \phi_{e'}})_{(VS)} +
(\frac{d^{2} \sigma}{d y d \phi_{e'}})_{(AT)} &=& B
|\mbox{\boldmath
$\eta_{\nu}^{' \perp}$}| \sqrt{\frac{m_{e}}{E_{\nu}}y[2-(2+\frac{m_{e}}{E_{\nu}})y]}\\
\times \{|g_{V}^{L}||g_{S}^{R}|cos(\phi-\alpha_{SV}) & + &
2|g_{A}^{L}||g_{T}^{R}|cos(\phi-\alpha_{TA})\}, \nonumber \eeq
where $\alpha_{SV} \equiv\alpha_{S}^{R} - \alpha_{V}^{L}$,
$\alpha_{TA} \equiv\alpha_{T}^{R} - \alpha_{A}^{L} $- the relative
phases between the  $g_{S}^{R}$, $g_{V}^{L}$ and $g_{T}^{R}$,
$g_{A}^{L}$ couplings respectively, $\phi$ - the angle between
the reaction plane and the transverse neutrino polarization
vector and is connected with the $\phi_{e'}$ in the following way;
$\phi=\phi_{0}-\phi_{e'}$, where $\phi_{0}$ - the angle between
the production plane and the transverse neutrino polarization
vector,  Fig. \ref{wsp}. \par
\begin{figure}
\includegraphics[%
  bb=80bp 20bp 540bp 0bp,
  width=6.5cm,
  angle=270]{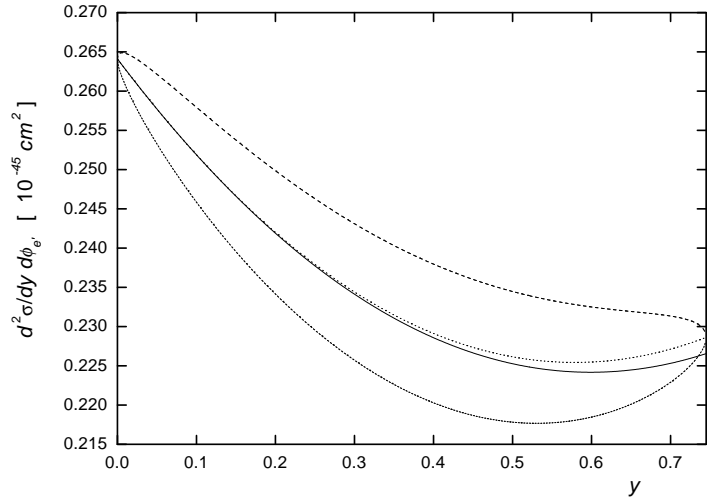}
\caption{Plot of the $\frac{d^{2} \sigma}{d y d \phi_{e'}}$ as a
function of $y$ for the $(\nu_{\mu}e^{-})$ scattering,
$E_{\nu}=0.746\, MeV$; a) SM with the left-handed neutrino (solid
line), b) the case of the exotic S, T, P  couplings  of the
right-handed neutrinos for $\phi-\alpha_{SV}=0,
\phi-\alpha_{TA}=0$ (long-dashed line), $\phi-\alpha_{SV}=\pi,
\phi-\alpha_{TA}=\pi$ (short-dashed line) and
$\phi-\alpha_{SV}=\pi/2, \phi-\alpha_{TA}=\pi/2$ (dotted line),
respectively.}
 \label{Wy1MM}
\end{figure}

 The presence of the interference
terms between the standard and exotic couplings in the cross
section depending on the $\phi_{e'}$ generates the azimuthal
asymmetry  in the angular distribution of the recoil electrons.
Because the right-handed neutrinos are produced and detected in
the exotic $(S, T, P)$ interactions, one uses the same upper
limits on the $g_{S}^{R}, g_{T}^{R}, g_{P}^{R}$ as for the
$C_{S}^{R}, C_{T}^{R}, C_{P}^{R}$, assuming the universality of
weak interactions. We take the values $|\mbox{\boldmath
$\eta_{\nu}^{' \perp}$}| = 0.414$, $\mbox{\boldmath
$\hat{\eta}_{\nu}$}\cdot\hat{\bf q}=-0.910$ for the
muon-neutrinos to get the upper limit on the expected effect of
the azimuthal asymmetry in the cross section.  The situation is
illustrated in the Fig. \ref{Wy1MM}. The plot for the SM is made
with the use of the present experimental values for
$g_{V}^{L}=-0.040\pm 0.015$, $g_{A}^{L}=-0.507\pm 0.014$
\cite{Data}, when $\mbox{\boldmath
$\hat{\eta}_{\nu}$}\cdot\hat{\bf q}= -1$, Fig. \ref{Wy1MM} (solid
line). If one integrates over the $\phi_{e'}$, both interference
terms vanish and the cross section $ d \sigma/d y$ consists of
only two terms.
\par If one assumes the production of only left-handed neutrinos in the standard
 $(V-A)$ and  non-standard $(S, T, P)$ weak interactions, there is no interference
between the  $g_{V, A}^{L}$ and $g_{S, T, P}^{L}$ couplings in
the differential cross section, when $m_{\nu}\rightarrow 0$, and
the azimuthal  distribution of the recoil electrons is symmetric.
We do not consider this scenario.
\par
\begin{figure}
\includegraphics[%
  bb=80bp 20bp 540bp 0bp,
  width=6.5cm,
  angle=270]{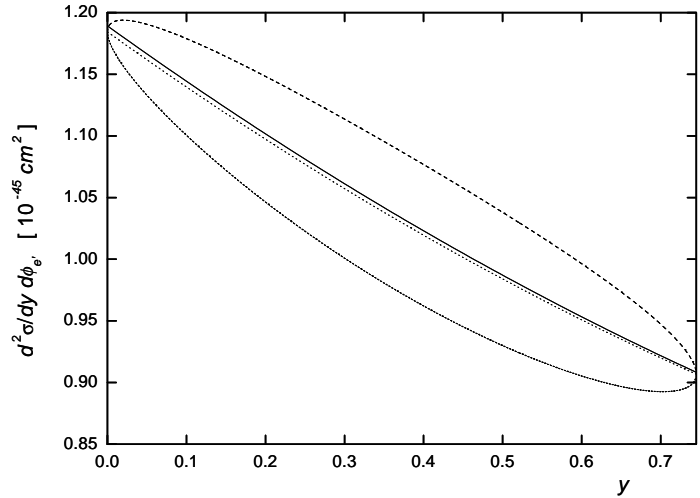}
\caption{Plot of the $\frac{d^{2} \sigma}{d y d \phi_{e'}}$ as a
function of $y$ for the $(\nu_{e}e^{-})$ scattering,
$E_{\nu}=0.746\, MeV$; a) SM with the left-handed neutrino (solid
line), b) the case of the exotic S, T, P  couplings  of the
right-handed neutrinos for $\phi-\alpha_{SV}=0,
\phi-\alpha_{TA}=0$ (long-dashed line), $\phi-\alpha_{SV}=\pi,
\phi-\alpha_{TA}=\pi$ (short-dashed line) and
$\phi-\alpha_{SA}=\pi/2, \phi-\alpha_{TA}=\pi/2$ (dotted line),
respectively.}
 \label{Wy3MM}
\end{figure}

 Considering  the low-energy intense artificial $(^{51}Cr)$ and
natural (Sun) electron-neutrino sources, we show the upper limit
on  the expected magnitude of the azimuthal asymmetry in the cross
section for the electron-neutrinos, Fig. \ref{Wy3MM}. It can be
noticed that the possible effect is much larger than for  the
muon-neutrinos at the same neutrino energy $E_{\nu}=0.746\, MeV$.
In the case of the $(\nu_{e} e^{-})$ scattering, the charged
current weak interaction must be included, i. e.  $g_{V}^{L}+1, \;
g_{A}^{L}+1$. We use the same upper limits on the exotic
couplings  as for the $g_{S}^{R}, g_{T}^{R}, g_{P}^{R}$, assuming
the universality of weak interactions. We also take
 the values $|\mbox{\boldmath $\eta_{\nu}^{' \perp}$}|= 0.414$,
$\mbox{\boldmath $\hat{\eta}_{\nu}$}\cdot\hat{\bf q}=-0.910$. The
plot for the SM is made with the same values of the standard
coupling constants as for the $(\nu_{\mu} e^{-})$ process, i. e.
$ -0.040 + 1$, $ -0.507 + 1$, when $\mbox{\boldmath
$\hat{\eta}_{\nu}$}\cdot\hat{\bf q}= -1$, Fig. \ref{Wy3MM} (solid
line).
\section{Astrophysical sources of right-handed \protect \\
neutrinos - neutrino spin flip} 
\begin{figure}
\includegraphics[%
  bb=80bp 10bp 540bp 0bp,
  width=6.5cm,
  angle=270]{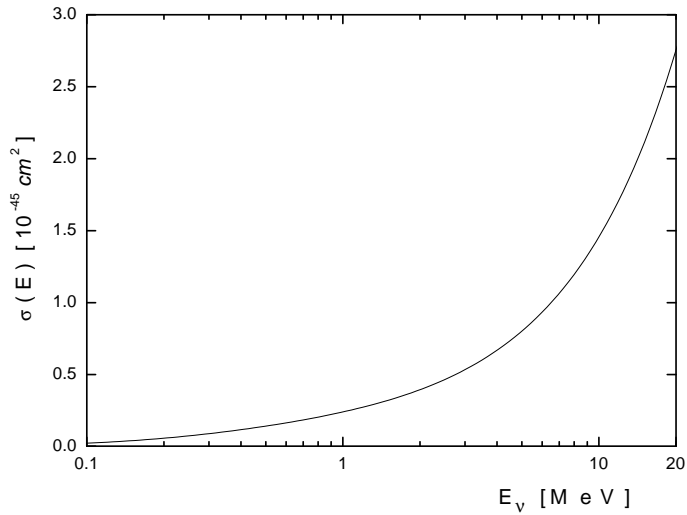}
\caption{Plot of the total cross section $\sigma(E)$ as a
function of right-handed $(\mbox{\boldmath
$\hat{\eta}_{\nu}$}\cdot\hat{\bf q}=1)$ neutrino energy $E_\nu$
for the $(\nu_e e^{-})$ scattering.} \label{Wy5MM}
\end{figure}

If a neutrino has a large
magnetic moment, the helicity of a neutrino can be flipped when
it passes through a region with magnetic field perpendicular to
the direction of propagation. The spin flip would change the
left-handed neutrino that is active in SM (V, A left-couplings)
into a right-handed neutrino $(\mbox{\boldmath
$\hat{\eta}_{\nu}$}\cdot\hat{\bf q}=1)$ that is sterile in SM:
\begin{eqnarray}
(\frac{d^{2}\sigma}{dyd\varphi})_{(V,A)}=\,(1-\mbox{\boldmath
$\hat{\eta}_{\nu}$}\cdot\hat{\bf q})\,\cdot\, f(E_{\nu}, y) &
=\,0.\label{eq:VA}
\end{eqnarray}
The mechanism of neutrino {}``spin flip'' in the Sun's convection
zone is proposed to explain the observed depletion of the solar
neutrinos \cite{key-1}. The most restrictable bound on the
neutrino magnetic moment arrives from astrophysical consideration
of a supernova explosion. The scattering due to the photon
exchange between a neutrino and a charged particle in plasma
leads to neutrino spin flip. The energy released in supernova
implosion is taken partly away by sterile neutrinos without
further interactions. In this scenario the neutrino magnetic
moment should be bounded because of the observed neutrino signal
of SN 1987A\cite{key-2}. Our paper shows that the participation
of the exotic couplings of the right-handed neutrinos can modify
the both astrophysical considerations. The right-handed neutrino
is no longer {}``sterile''. The total cross section for
$\nu_{e}e^{-}$ scattering with the coupling constants from the
current data (Section 2) can be calculated from our general
formulas (see Fig. \ref{Wy5MM}). In this scenario the
right-handed neutrinos can be detected by neutrino detectors and
could help simultaneously to transfer the energy to presupernova
envelope.
\par
\begin{figure}
\includegraphics[%
  bb=80bp 20bp 540bp 0bp,
  width=6.5cm,
  angle=270]{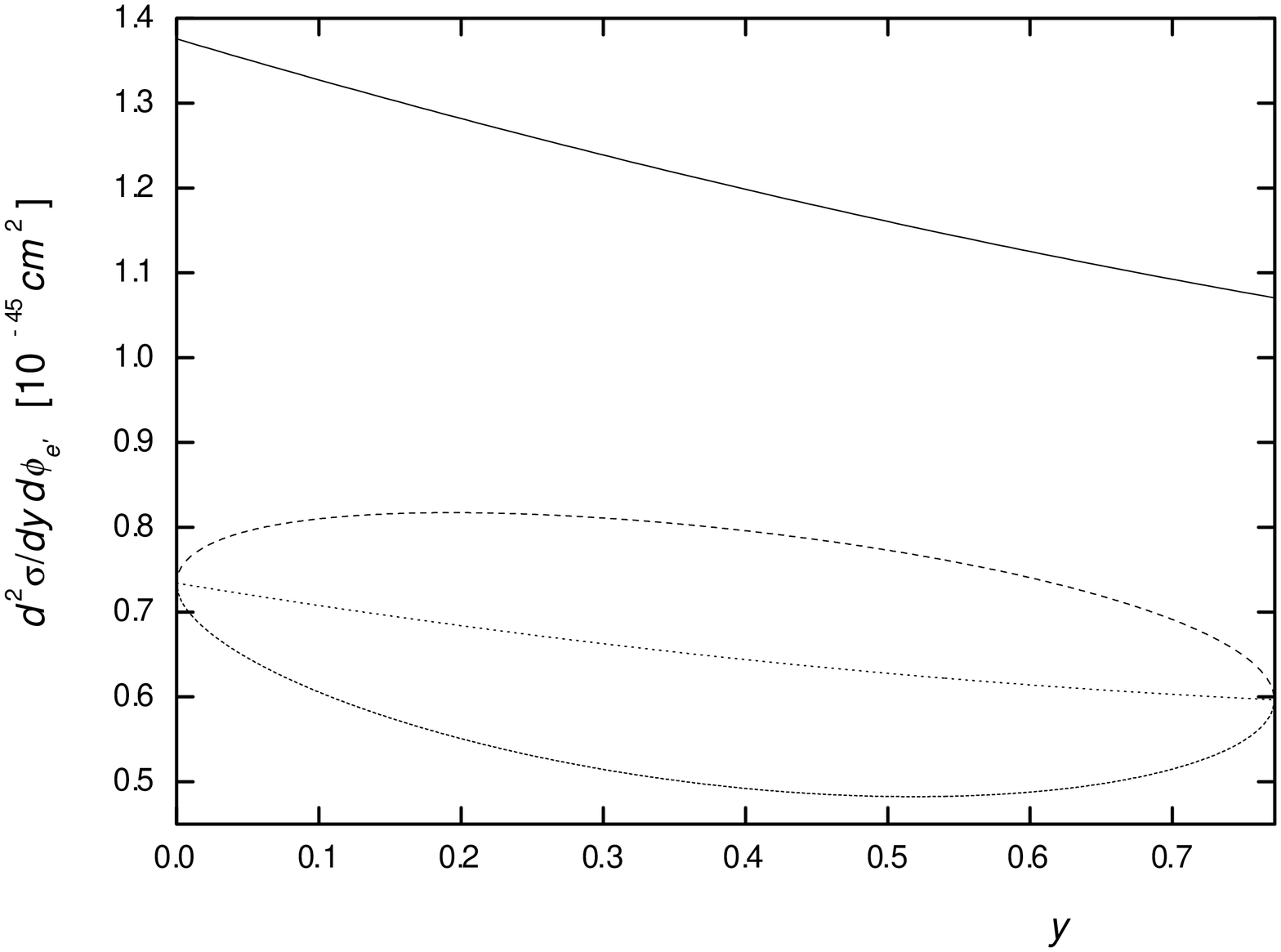}
\caption{Plot of the $\frac{d^{2} \sigma}{d y d \phi_{e'}}$ as a
function of $y$ for the $(\nu_{e}e^{-})$ scattering,
$E_{\nu}=0.863\, MeV$; a) SM with the left-handed neutrino (solid
line) for $ \mbox{\boldmath $\hat{\eta}_{\nu}$}\cdot\hat{\bf q}=
-1$, b) the case of the exotic S, T, P couplings of the
right-handed neutrinos for $\phi-\alpha_{SV}=0,
\phi-\alpha_{TA}=0$ (long-dashed line), $\phi-\alpha_{SV}=\pi,
\phi-\alpha_{TA}=\pi$ (short-dashed line) and
$\phi-\alpha_{SA}=\pi/2, \phi-\alpha_{TA}=\pi/2$ (dotted line),
respectively, when $\mbox{\boldmath
$\hat{\eta}_{\nu}$}\cdot\hat{\bf q} = 0$, $|\mbox{\boldmath
$\eta_{\nu}^{' \perp}$}|=1$.}
 \label{Wy6MM}
\end{figure}

 If the conversions $\nu_{eL}\rightarrow \nu_{eR}$ in the Sun
are possible, the azimuthal asymmetry in the angular distribution
of the recoil electrons generated by the interference terms
between the standard $(V, A)_{L}$ and exotic $(S, T, P)_{R}$
couplings should occur. If one assumes that a survival probability
 for the left-handed $^{7}Be$-neutrinos  is equal to
 $P_{eL}=0.5$,
the value of the transverse neutrino polarization as a function of
this $P_{eL}$ will be large, $|\mbox{\boldmath $\eta_{\nu}^{'
\perp}$}|=2\sqrt{P_{eL}(1-P_{eL})}=1$, (for this case
$\mbox{\boldmath $\hat{\eta}_{\nu}$}\cdot\hat{\bf q}= 1-2\cdot
P_{eL}=0$), see Eq. (9) in \cite{Barbieri}. The equation on the
$|\mbox{\boldmath $\eta_{\nu}^{' \perp}$}|$ arises from the
density matrix for the relativistic neutrino chirality.
 The situation is illustrated in the Fig. \ref{Wy6MM} for the same limits on the exotic
 couplings as for the $^{51}Cr$-neutrinos and $E_{\nu}=0.863\,MeV$. In this
way, the expected effect of the azimuthal asymmetry would be much
 stronger than for the $^{51}Cr$-neutrinos.
\section{Conclusions}
In this paper, we show that the production of the R-handed
neutrinos in the exotic $(S, T, P)$ weak interactions in addition
to the L-handed ones in the standard $V-A$ weak interaction should
manifest in the observation of the azimuthal asymmetry of the
electrons recoiled  after the subsequent  neutrino scattering.
This asymmetry would be due to the terms with the interference
between  $(V, A)_{L}$  and $(S, T, P)_{R}$ weak interactions,
which stay present even in the limit of massless neutrino.  The
scenario with interfering L- and R-handed neutrinos could be
tested with the intense electron-neutrino beams, e.g. from the
artificial polarized $^{51}Cr$ source, Fig. \ref{Wy3MM}. If the
neutrino helicity flip in the solar magnetic field takes place,
the similar effect of the azimuthal asymmetry in the event number
for the solar neutrinos should appear,  Fig. \ref{Wy6MM}
($^{7}Be$-neutrinos). It would indicate the neutrino spin flip
scenario $(\nu_{L} \rightarrow \nu_{R})$ as a possible solution
of the observed solar neutrino deficit. In both cases, the
azimuthal asymmetry would arise from the interference terms
between the standard $(V, A)_{L}$ and $(S, T, P )_{R}$ exotic
couplings, proportional to the transverse neutrino polarization
vector.
\par It is well-known that according to the SM the angular
distribution of the recoil electrons does not depend on the
azimuthal angle $\phi_{e'}$, i.e. is the azimuthally symmetric.
The detection of the azimuthal asymmetry would be a signature of
the R-handed neutrinos.
 It can be noticed that the expected effect would be much stronger for
 the low-energy
neutrino-electron scattering $(E_{\nu}< 1 \, MeV )$ than for the
high-energy one. The artificial neutrino source has to be
polarized to have the assigned direction of the transverse
neutrino polarization vector with respect to the production plane
because it would allow to measure the $\phi_{e'}$. In the case of
the solar neutrinos, the $\mbox{\boldmath $\eta_{\nu}^{' \perp}$}$
would be directed along the solar magnetic field. The neutrino
detectors with the good angular resolution have to observe the
direction of the recoil electrons and to analyze all the possible
reaction planes corresponding to the given recoil electron
scattering angle in order to verify if the azimuthal asymmetry in
the cross section appears.

\vspace{.5cm} This work was supported in part by the grant 2P03B
15522 of The Polish Committee for Scientific Research and by The
Foundation for Polish Science.

\section{Appendix}
 The formulas for the 4-vector initial neutrino polarization $S$ in its rest
 frame and for the initial neutrino moving  with the momentum ${\bf q}$,
 respectively, are as follows:
\beq S & = & (0,\mbox{\boldmath $\hat{\eta}_{\nu}$}),\\
 S' & = & \frac{\mbox{\boldmath $\hat{\eta}_{\nu}$}\cdot{\bf q}}{E_{\nu}}\cdot
\frac{1}{m_{\nu}} \left(
\begin{array}{c}  E_{\nu}\\ {\bf q} \end{array} \right)
 + \left(
\begin{array}{c}  0\\ \mbox{\boldmath $\hat{\eta}_{\nu}$}  \end{array} \right) -
\frac{\mbox{\boldmath $\hat{\eta}_{\nu}$}\cdot{\bf
q}}{E_{\nu}(E_{\nu}+m_{\nu})} \left( \begin{array}{c}  0\\ {\bf q}
\end{array} \right), \\
S^{0'} & = & \frac{{|\bf q|}}{m_{\nu}}(\mbox{\boldmath
$\hat{\eta}_{\nu}$}\cdot{\bf \hat{q}}), \\
{\bf S'} & = & \frac{E_{\nu}}{m_{\nu}}(\mbox{\boldmath
$\hat{\eta}_{\nu}$}\cdot{\bf \hat{q}}){\bf \hat{q}} +
\mbox{\boldmath $\hat{\eta}_{\nu}$} - (\mbox{\boldmath
$\hat{\eta}_{\nu}$}\cdot{\bf \hat{q}}){\bf \hat{q}},
 \eeq
 where $\mbox{\boldmath $\hat{\eta}_{\nu}$}$ - the unit vector of the
 initial neutrino polarization in its rest frame.
 The formulas for the  Michel-Wightman density matrix \cite{Michel} in the
 case of the polarized neutrino with the non-zero mass,
 \beq
\Lambda_{\nu}^{(s)}& = & \sum_{r=1, 2}u_{r}\overline{u}_{r} \sim
[1+\gamma_{5}(S^{'\mu}\gamma_{\mu})][(q^{\mu}\gamma_{\mu}) +
m_{\nu}] \\
& = & \mbox{} [(q^{\mu}\gamma_{\mu}) + m_{\nu} +
\gamma_{5}(S^{'\mu}\gamma_{\mu})(q^{\mu}\gamma_{\mu}) +
\gamma_{5}(S^{'\mu}\gamma_{\mu}) m_{\nu}],  \nonumber\\
(S^{'\mu}\gamma_{\mu}) &=& \frac{\mbox{\boldmath
$\hat{\eta}_{\nu}$}\cdot{\bf q}}{E_{\nu}m_{\nu}}(q^{\mu} \gamma_{\mu}) -
(\mbox{\boldmath $\hat{\eta}_{\nu}$} - \frac{(\mbox{\boldmath
$\hat{\eta}_{\nu}$}\cdot{\bf q}){\bf q}}{E_{\nu}(E_{\nu}+
m_{\nu})})\cdot\mbox{\boldmath$\gamma$}, \\
(S^{'\mu}\gamma_{\mu})(q^{\mu}\gamma_{\mu}) &=&
\frac{m_{\nu}}{E_{\nu}}\mbox{\boldmath $\hat{\eta}_{\nu}$}\cdot{\bf q} -
(\mbox{\boldmath $\hat{\eta}_{\nu}$} - \frac{(\mbox{\boldmath
$\hat{\eta}_{\nu}$}\cdot{\bf q}){\bf q}}{E_{\nu}(E_{\nu}+
m_{\nu})})\cdot \mbox{\boldmath $\gamma$}(q^{\mu}\gamma_{\mu}),
\\ (S^{'\mu}\gamma_{\mu}) m_{\nu} &=& \frac{\mbox{\boldmath
$\hat{\eta}_{\nu}$}\cdot{\bf q}}{E_{\nu}}(q^{\mu}\gamma_{\mu}) -
m_{\nu}(\mbox{\boldmath $\hat{\eta}_{\nu}$} -
\frac{(\mbox{\boldmath $\hat{\eta}_{\nu}$}\cdot{\bf q}){\bf
q}}{E_{\nu}(E_{\nu}+ m_{\nu})})\cdot \mbox{\boldmath $\gamma$},
\eeq and in the limit of vanishing
 neutrino mass, we have
\beq \lim_{m_{\nu}\rightarrow 0} \Lambda_{\nu}^{(s)}
 & = & \mbox{} [1 + \gamma_{5}[\frac{\mbox{\boldmath
$\hat{\eta}_{\nu}$} \cdot{\bf q}}{|{\bf q}|} - (\mbox{\boldmath
$\hat{\eta}_{\nu}$} - \frac{(\mbox{\boldmath
$\hat{\eta}_{\nu}$}\cdot{\bf q}){\bf q}}{|{\bf q}|^{2}})\cdot
\mbox{\boldmath $\gamma$}]](q^{\mu}\gamma_{\mu}).  \eeq We see
that in spite of the singularities $m_{\nu}^{-1}$ in the
longitudinal component $(\mbox{\boldmath
$\hat{\eta}_{\nu}$}\cdot{\bf \hat{q}})$, the Michel-Wightman
density matrix, in the limit of vanishing neutrino mass $m_{\nu}$,
remains finite including the transverse component of neutrino
polarization.

\end{document}